\documentclass[conference]{IEEEtran}
  {\list{}{\leftmargin=0.07in\rightmargin=0.07in}\item[]}%
  {\endlist}

\usepackage{multirow}
\usepackage[flushleft]{threeparttable}

\usepackage{tikz}
\usepackage{amsmath}

\usepackage{graphicx}
\graphicspath{{Images/}}

\usepackage{hyperref}

\usepackage{subfigure}

\usepackage{microtype}
\def\ie{{i.e.},~}
\def\eg{{e.g.},~}

\usepackage{pifont}
\newcommand{\cmark}{\ding{51}}%
\newcommand{\xmark}{\ding{55}}%

\usepackage{xcolor}

\usepackage{booktabs}
\usepackage{setspace}

\usepackage{amsmath,amssymb}
\usepackage[ruled,vlined,linesnumbered]{algorithm2e}

\newcommand{\berkay}[1]{\textcolor{red}{[Berkay: #1]}}
\newcommand{\mikey}[1]{\textcolor{blue}{[Mikey: #1]}}

\usepackage{array}
\newcolumntype{P}[1]{>{\centering\arraybackslash}p{#1}}

\DeclareRobustCommand*\circled[1]{\tikz[baseline=(char.base)]{ \node[shape=circle,draw,color=white,fill=black,inner sep=0.5pt] (char){#1};}}

\usepackage{comment}

\ifCLASSINFOpdf
\else
\fi

\makeatletter
\newcommand*\titleheader[1]{\gdef\@titleheader{#1}}
\AtBeginDocument{%
  \let\st@red@title\@title%
  \def\@title{%
    \bgroup\normalfont\large\centering\@titleheader\par\egroup
    \vskip0.2em\st@red@title}
}
\makeatother

\title{On the Safety Implications of Misordered Events and Commands in IoT Systems}
\titleheader{Accepted to the IEEE Workshop on the Internet of Safe Things (SafeThings), 2021.}

\begin{document}

%







\author{\IEEEauthorblockN{Furkan Goksel\IEEEauthorrefmark{1}\textsuperscript{\textsection}\textsuperscript{\textparagraph}, Muslum Ozgur Ozmen\IEEEauthorrefmark{2}\textsuperscript{\textsection}, Michael Reeves\IEEEauthorrefmark{2}, Basavesh Shivakumar\IEEEauthorrefmark{2} and Z. Berkay Celik\IEEEauthorrefmark{2}}\IEEEauthorblockA{\IEEEauthorrefmark{1}Middle East Technical University, furkan.goksel@metu.edu.tr}\IEEEauthorblockA{\IEEEauthorrefmark{2}Purdue University, \{mozmen, reeves17, bammanag, zcelik\}@purdue.edu}}

\maketitle
\begingroup\renewcommand\thefootnote{\textsection}
\footnotetext{contributed equally.}
\endgroup
\begingroup\renewcommand\thefootnote{\textparagraph}
\footnotetext{All works were done during this author's research internship at Purdue University.}
\endgroup

\thispagestyle{plain}
\pagestyle{plain}

\begin{abstract}
IoT devices, equipped with embedded actuators and sensors, provide custom automation in the form of IoT apps. 
IoT apps subscribe to events and upon receipt, transmit actuation commands which trigger a set of actuators. 
Events and actuation commands follow paths in the IoT ecosystem such as sensor-to-edge, edge-to-cloud, and cloud-to-actuator, with different network and processing delays between these connections.
Significant delays may occur especially when an IoT system cloud interacts with other clouds.
%
Due to this variation in delays, the cloud may receive events in an incorrect order, and in turn, devices may receive and actuate misordered commands.
In this paper, we first study eight major IoT platforms and show that they do not make strong guarantees on event orderings to address these issues.
We then analyze the end-to-end interactions among IoT components, from the creation of an event to the invocation of a command.
From this, we identify and formalize the root causes of misorderings in events and commands leading to undesired states.
We deploy 23 apps in a simulated smart home containing 35 IoT devices to evaluate the misordering problem.
Our experiments demonstrate a high number of misordered events and commands that occur through different interaction paths.
Through this effort, we reveal the root and extent of the misordering problem and guide future work to ensure correct ordering in IoT systems.

\end{abstract}

\section{Introduction}\label{sec:introduction}
Commodity IoT systems are mainly composed of sensors, actuators, edge devices, and clouds.
Actuators enact changes in the physical space through commands, and sensors report changes in the physical space through events~\cite{ozmen2021discovering}.
The edge forms a bridge between the cloud and IoT devices translating local wireless communication (e.g. zigbee and bluetooth) to traditional packet switched networks.
The cloud maintains the state of the IoT devices and provides interfaces for automated control of actuators through IoT apps.
IoT apps are simple programs used to create automation. 
An app subscribes to sensor events (e.g. motion-active) or other common events (e.g. voice or button pushes), and actuates a set of devices when an event is received~\cite{soteria}.
For instance, an app locks the door when a user taps an icon through their mobile device.

\begin{figure}[t!]
    \centering
    \includegraphics[width=0.95\columnwidth]{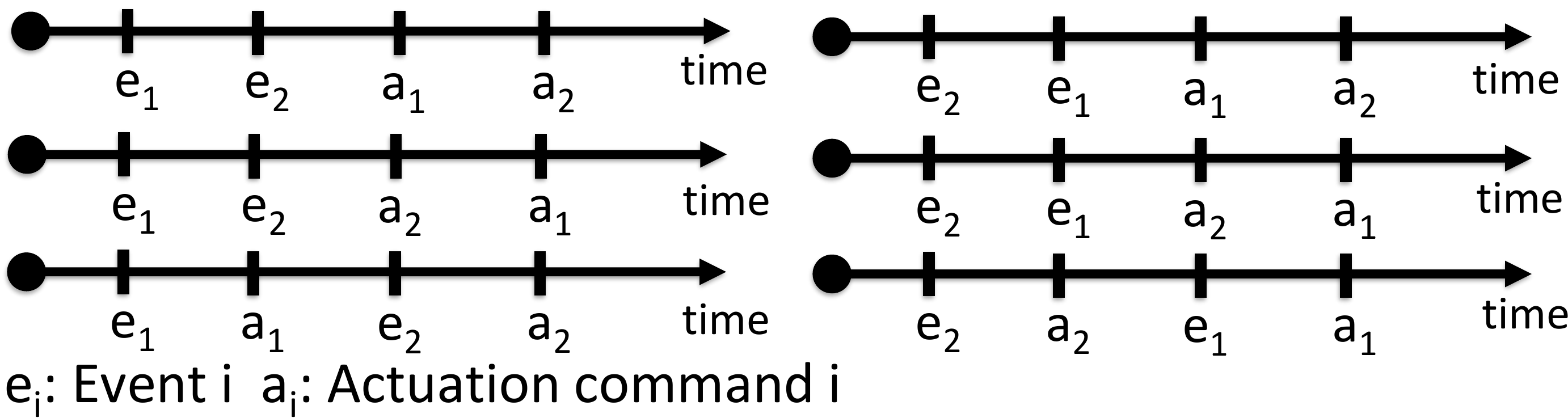}
    \caption{Possible arrival time of events ($\mathtt{e_1, e_2}$) to the cloud and actuation commands to the actuators ($\mathtt{a_1, a_2}$) (An app subscribed to $\mathtt{e_1}$ actuates $\mathtt{a_1}$, and another app subscribed to $\mathtt{e_2}$ actuates $\mathtt{a_2}$).}
    \label{fig:problem}
\end{figure}

\vspace{1pt}\noindent\textbf{Misordering Problem.} Events and actuation commands have different processing time and network latency while traversing diverse IoT components.
For instance, in its simplest form, a sensor transmits an event to the edge, the edge then transmits it to the cloud. The cloud processes the event to find the apps subscribed to it, obtains the actuation commands by checking apps' event handler methods, and sends the actuation commands to the edge. Lastly, the edge sends the commands to the actuators.
The "misordering" problem arises when the events arrive at the cloud in different orders, which further causes devices to actuate out of order. 
Even if events arrive at the cloud in the correct order, corresponding actuation commands may still arrive at the actuators in a wrong order. 
To illustrate, the cloud receives events from two different sensors ($\mathtt{e_1}$ and $\mathtt{e_2}$).
It then sends two separate actuation commands to the devices that two apps subscribed to these events ($\mathtt{app_1}$ $\rightarrow$ $\mathtt{a_1}$ and $\mathtt{app_2}$ $\rightarrow$ $\mathtt{a_2}$).
The only guarantee provided by an IoT system is $\mathtt{e_1}$ must happen before $\mathtt{a_1}$ and $\mathtt{e_2}$ must happen before $\mathtt{a_2}$ since the events trigger the actuation commands.
This guarantee leads to six possible sequences that consist of different event and actuation command orders, as illustrated in Figure~\ref{fig:problem}. 

\begin{table*}[th!]
\caption{The event and actuator command ordering guarantee provided by IoT programming platforms.}
\renewcommand{\arraystretch}{1}   
\resizebox{\textwidth}{!}{%
\begin{threeparttable}[b]{
\begin{tabular}{|c|c|c|c|c|}
\hline
\multicolumn{2}{|c|}{\textbf{IoT Platform}} &
  \textbf{\begin{tabular}[c]{@{}c@{}}Order \\ Guarantee\end{tabular}} &
  \textbf{\begin{tabular}[c]{@{}c@{}}Developer \\ Advice\end{tabular}} &
  \textbf{Details} \\ \hline
\multirow{3}{*}{\textbf{General}} &
  Amazon IoT &
  \xmark &
  \cmark &
  ``The accuracy of the timestamp   that event occurred is +/- 2 minutes...''~\cite{amazonEventGuarantee} \\ \cline{2-5} 
 &
  Google IoT &
  \xmark &
  \cmark &
  ``The order in which messages   are received by subscribers is not guaranteed...''~\cite{google} \\ \cline{2-5} 
 &
  Microsoft IoT &
  \xmark &
  \cmark &
  ``The Azure event grid does not support ordering of events...''~\cite{microsoftevent} \\ \hline
\multirow{3}{*}{\textbf{SmartHome}} &
  OpenHAB &
  \xmark &
  \cmark &
  \begin{tabular}[c]{@{}c@{}}There is no mention on order event   guarantees~\cite{openhabevent}, however advice is present~\cite{openhabpost}\end{tabular} \\ \cline{2-5} 
 &
  SmartThings &
  \xmark &
  \xmark &
  \begin{tabular}[c]{@{}c@{}}``The platform follows eventually consistent programming, meaning that responses  to a request for \\ a value in IoT apps will eventually be the same, but in the short term they might differ...''~\cite{smartthingsevents}\end{tabular} \\ \hline
\multirow{5}{*}{\textbf{Trigger-Action}} &
  IFTTT &
  \xmark &
  \xmark &
  ``Action conflicts resulting in non-deterministic event orderings were detected in previous work...''~\cite{trig-iftt} \\ \cline{2-5} 
 &
  Microsoft Power Automate &
  \xmark &
  \cmark &
  \begin{tabular}[c]{@{}c@{}}  "Microsoft Power Automate implements a simple FIFO queue so that \\ only a single event executes in parallel, but does not handle out of order events."~\cite{flowevent}\end{tabular} \\  \cline{2-5} 
  &
  Zappier &
  \xmark &
  \cmark &
  \begin{tabular}[c]{@{}c@{}}
  "Zapier does not guarantee order on race conditions: In the best case \\ the last event run will error, but you could also execute duplicate events"~\cite{zapierevent} \end{tabular}\\ \hline
\end{tabular}
}
\end{threeparttable}
}

\label{table:platformStudy}
\end{table*}

This observation causes two possible issues: $(1)$ undesired device states, and $(2)$ logging the events incorrectly.
First, the misordered events invoke actuation commands in an incorrect order, causing undesired final system states.
For example, a user gives a ``turn oven on'' ($\mathtt{a_1}$) command through a voice-enabled device ($\mathtt{e_1}$).
After a short period of time, the user gives another voice command ($\mathtt{e_2}$), ``turn oven off'' ($\mathtt{a_2}$). 
The user's desired state is where $\mathtt{a_1}$ happens before $\mathtt{a_2}$ so that the oven is turned-off. 
Yet, $\mathtt{e_2}$ arrives to the cloud before $\mathtt{e_1}$ and the commands arrive at the oven in the wrong order.
The oven's end state becomes $\mathtt{on}$, which is not desired or expected.
Second, the misordered IoT device states logged in the cloud (a) misguide users and (b) poison learning systems. 
To illustrate the first case, a user sends an ``unlock-door'' command through a mobile device.
Once it is executed, the door reports its ``door-unlocked'' state to the cloud. 
Soon after, the user sends a ``lock-door'' command and upon completion, the door similarly reports a ``door-locked'' state.
The ``door-locked'' and ``door-unlocked'' states  may not arrive at the cloud in the right order, leaving the door's end state on the cloud as ``unlocked''. 
For the second case, there are systems that leverage correlations among sensor and actuator states to perform specific tasks, such as activity recognition~\cite{han2018smart}, physical event verification~\cite{birnbach2019peeves}, IoT device pairing~\cite{mirzadeh2013secure,han2018you}, and anomaly detection~\cite{ghaeini2018state,ghaeini2019patt,sikder2019aegis}.
The misordered device states logged at the cloud may naturally poison the system logs, and a model trained on misordered logs may yield inaccurate results. 

\vspace{.5pt}\noindent\textbf{Challenges of the Misordering Problem in IoT.} 
Although event ordering is well-studied in wireless sensor networks and distributed systems~\cite{temporalmessage,distributed,onishi2020recovery}, IoT introduces unique challenges that hinder the seamless adoption of existing techniques. 
First, merely ensuring the event order does not guarantee the correct order of actuation commands in IoT deployments (See Figure~\ref{fig:problem}), possibly leaving the system in unsafe states.
Second, IoT systems often interact with \textit{device-vendor} and \textit{trigger-action} clouds to connect different services together.
To detail, an IoT cloud sends a request to a specific device-vendor cloud (\eg Philips) to actuate devices. 
For instance, a voice-command directing to turn off the lights first arrives at the cloud, which converts speech into text to identify the user's intent. 
It then transmits the turn-off intent to the device-vendor cloud, which directly sends the turn-off command to the lights. 
As another example, a user uses a trigger-action rule to connect the IoT platform with another service, \eg a user installs a trigger-action rule to log power meter states to a Google spreadsheet file~\cite{celik2019verifying}.
In these cases, the processing time and network latency between different system components become unpredictable, which increases the number of incorrectly ordered events and actuation commands.

\vspace{1pt}\noindent\textbf{Our Contributions.} In this paper, we analyze the event and command misordering problem in centralized IoT systems. 
First, we present our study on the official documentations of eight popular IoT programming platforms to identify how they handle the misordering problem. 
We then study the interactions among IoT components and apps to analyze the possible communication paths an event may take to result in an actuation command's execution. 
Through this, we identify and formalize the root causes of incorrect order of events and actuation commands leading to undesired states. 
Specifically, we take into account single or multiple actuators being invoked by single or multiple event sources, as well as the temporal and logical relations among different actuators.
Lastly, we evaluate the extent of the misordering problem in a simulated smart home containing $15$ different actuators and $7$ sensors, as a total of $35$ IoT devices, automated by $23$ apps.
The apps are triggered by different events such as voice or mobile applications, and they interact with different clouds. Over three different experiments, we set delays between IoT components collected from real world data and simulate the apps to identify undesired device states. 
We found through these experiments that on average, $13.5\%$ of events are received in an incorrect order by the user IoT cloud, and $29.8\%$ of actuation commands are received in an incorrect order by actuators.

\section{IoT Platform Study}
\label{sec:platform}
We present a study of eight popular IoT programming platforms, three general-purpose (Amazon IoT~\cite{amazonIoT}, Google IoT~\cite{googleIoT}, and Microsoft IoT~\cite{microsoftIoT}), two smart home (OpenHAB~\cite{openHab}, and SmartThings~\cite{smartThings}), and three trigger-action (IFTTT~\cite{IFTTT}, Microsoft Power Automate~\cite{powerAutomate}, and Zapier~\cite{Zapier}).
We studied their official documentations to identify whether they guarantee event ordering and provide developers advice to solve the misordering problem in their system design. 
The most recent results are presented in Table~\ref{table:platformStudy}.
We found that the ordering guarantee is on a best-effort basis, meaning that none of the IoT platforms guarantee ordered event delivery. For instance, Amazon IoT documentation says that \emph{ accuracy of the timestamp that an event occurred is $\pm$2 minutes}~\cite{amazonEventGuarantee}.
Some platforms, however, guide developers to address the misordering problem at the application level. For instance, Microsoft IoT suggests using a monotonically increasing sequence number to ensure the latest event is processed from a single event source~\cite{microsoftevent}.
However, their statement does not detail how to integrate a unique sequence number for each device, does not guarantee the order of actuation commands from multiple sources, and does not consider the interactions among different IoT system components.
Lastly, some developers suggest workarounds for the misordering problem in official community forums. 
For instance, an OpenHAB developer states, \emph{write some rules to cache up the [events] for a hundred msec or so, use the timestamp from persistence to determine the actual order they were sent and process them in order}~\cite{openhabpost}.

\section{IoT System Analysis}
\label{sec:analysis}
We present the interactions among IoT components (Section~\ref{sec:componentInteraction}), introduce the root causes of the undesired states due to misordered events and commands (Section~\ref{sec:misorder}), and formalize the identified root causes (Section~\ref{sec:formalize}).

\subsection{System Component Interactions}
\label{sec:componentInteraction}
\begin{figure}[t!]
    \centering
    \includegraphics[width=1\columnwidth]{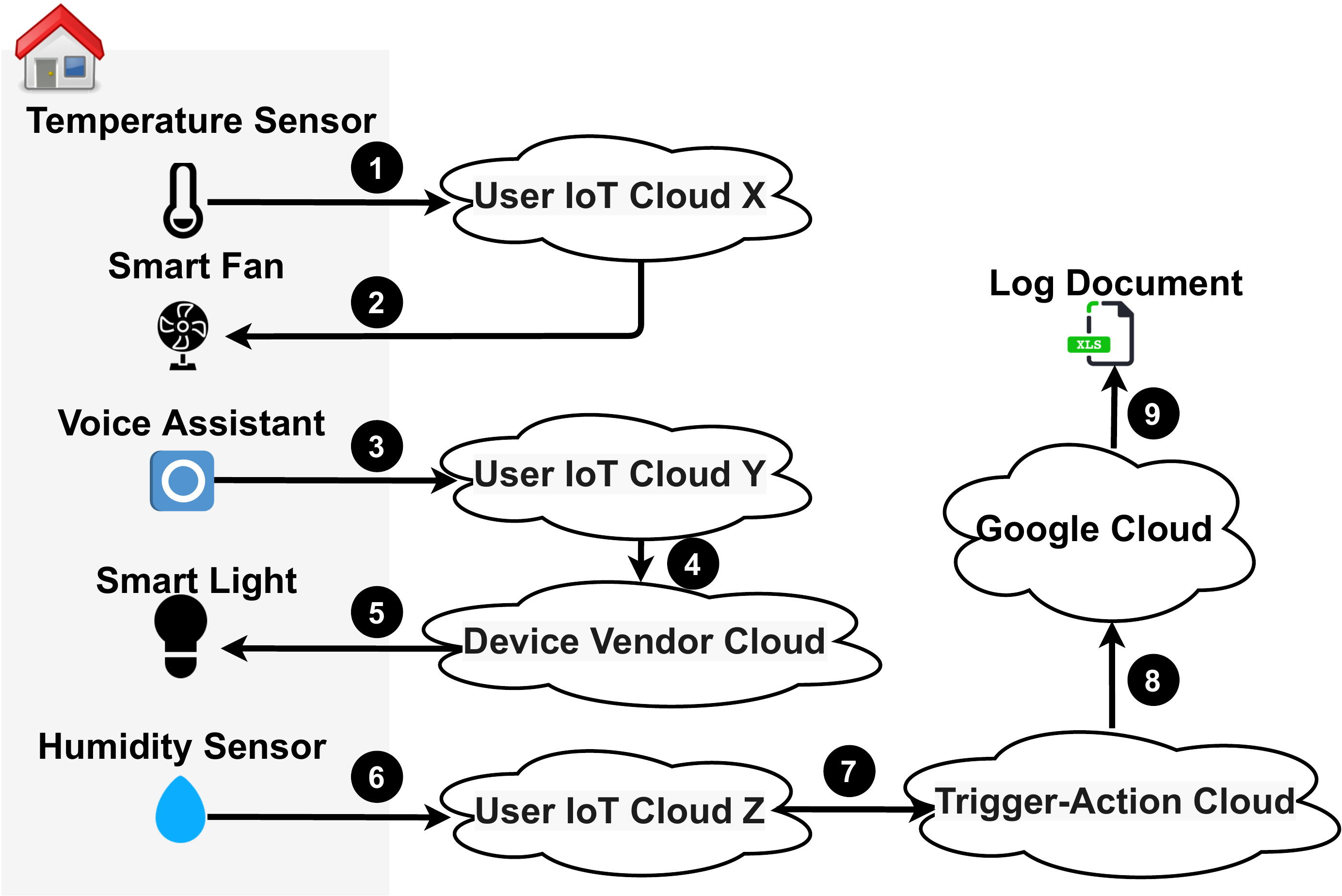}
    \caption{Illustration of different network paths taken by events and actuation commands in an IoT system.}
    \label{fig:systems}
\end{figure}

An actuation command can be invoked in an IoT system in three different ways: 
\textbf{(1)} A sensor event generated due to an environmental change is sent to an edge device. 
\textbf{(2)} A user sends an event via a mobile or desktop app provided by an IoT programming platform. 
\textbf{(3)} A user creates a ``voice'' event through a voice-enabled device. 
When the user IoT cloud receives an event from these sources, it first logs the event with the local cloud time. 
It then checks whether an app or a set of apps is subscribed to the event and determines if device actuation is required.
If the cloud determines no actuation is required, the operation terminates within the user IoT cloud. If actuations are required, there are often three possible interaction paths, as illustrated in Figure~\ref{fig:systems}, and detailed below.

\label{sect:analysis}

\vspace{1pt}\noindent\textbf{Case 1-} The user IoT cloud generates actuation commands within itself, meaning it does not interact with other clouds.
The cloud obtains the actuators by checking the apps' event handler methods to find the apps subscribed to the event.  
It then transmits an actuation command to the actuator through the edge device. 
For instance, a temperature sensor reports its measurements to the cloud through an edge device (\circled{$1$}). The user IoT cloud then identifies an app subscribed to the temperature changes, which turns on a smart fan. Therefore, the cloud sends the ``smart-fan-on'' command to the edge, and the edge transmits it to the smart fan (\circled{$2$}).

\vspace{1pt}\noindent\textbf{Case 2-} The user IoT cloud sends a request to the device vendor cloud to handle the actuation command requests.
For this, the user IoT cloud forwards the actuation command to a specific device-vendor cloud, which directly issues the actuation commands to the devices. This design is often used to delegate device management to the vendors and, in some cases, device-vendors desire to control their proprietary devices~\cite{amazonIoT}. 
For instance, a user gives the ``turn on light'' command to the voice assistant, which transmits the command to the user IoT cloud (\circled{3}). The user IoT cloud converts the voice to speech and identifies the user's intent. It then sends the command to the device vendor cloud  (\eg Philips Hue) (\circled{4}), which directly communicates to the light device to turn it on (\circled{5}).

\vspace{1pt}\noindent\textbf{Case 3-} Users often authorize third-party services such as trigger-action platforms (\eg IFTTT and Zapier) to connect different services and for data visualization and energy management. 
In these cases, the user IoT cloud transmits an event to the trigger action cloud when a specific event arrives, or the trigger-action cloud periodically polls the user IoT cloud to identify events.
The trigger-action cloud then checks whether a user has a rule subscribed to that event. It then sends an actuation command to the other vendor clouds or user IoT cloud based on the actuation.
To illustrate, consider a trigger-action rule that logs the humidity sensor readings to the Google spreadsheet. The humidity sensor measurements are first transmitted to the user IoT cloud (\circled{6}), and the user IoT cloud then transmits the measurements to the trigger-action cloud (\circled{7}). Trigger-action cloud sends a request to the Google Cloud (\circled{8}), which saves the humidity measurement to the spreadsheet file (\circled{9}).

\begin{figure*}[t!]
    \centering
    \includegraphics[width=1\textwidth]{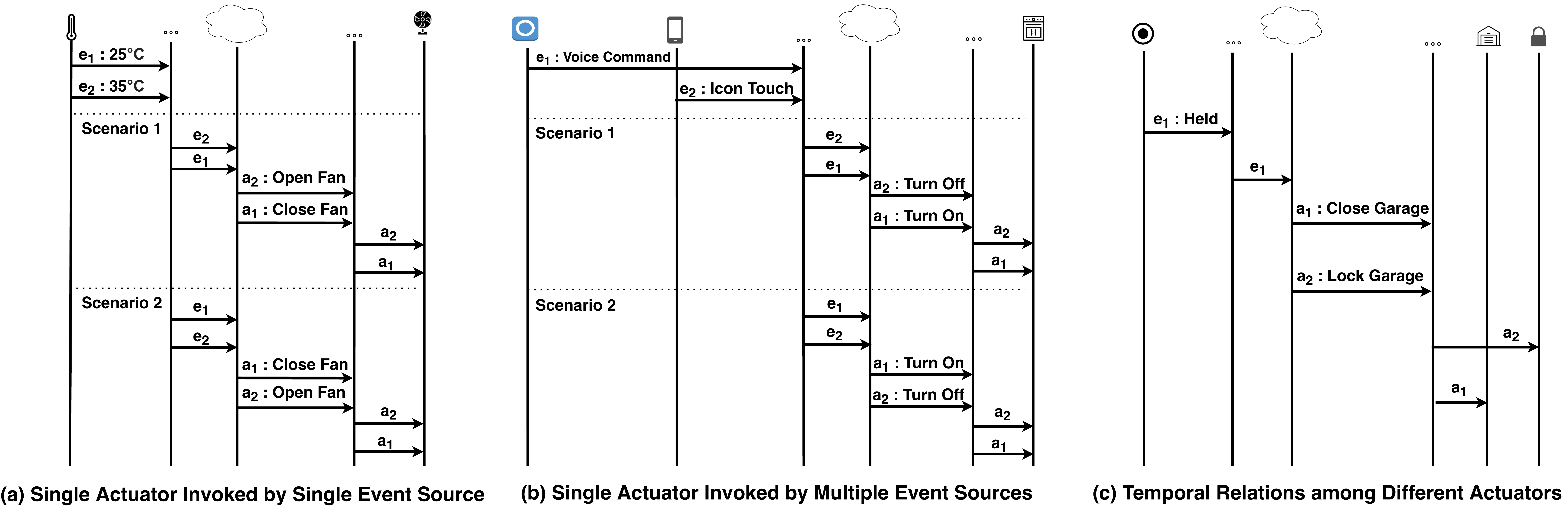}
    \caption{Illustration of root causes of undesired system states.}
    \label{fig:allproblems}
\end{figure*}

\subsection{Event and Command Misordering}
\label{sec:misorder}
Sensor events may arrive at the cloud in a different order than they were created.
This causes the cloud to issue actuation commands in the wrong order, leading to different device states other than the desired ones. 
Even if the sensor events arrive at the cloud correctly, the commands may still arrive at the actuators in the wrong order.
Below, we identify the root causes of incorrect event and command orders leading to undesired states by studying their relations with IoT apps.

\vspace{1pt}\noindent\textbf{Single Actuator Invoked by a Single Event Source.} 
We analyze multiple actuation commands issued to an actuator through a single event source.
When the actuation commands are the same, multiple commands are idempotent, resulting in the desired system state.
For instance, multiple ``oven-on'' commands sent through a mobile app leave the oven in an ``on'' state regardless of their order. 
Yet, if the commands are different, undesired system states may occur. 
To illustrate, consider an app that turns off a smart fan when the temperature is $\leq \mathtt{30^{\circ}C}$ and turns it on when the temperature is $> \mathtt{30^{\circ}C}$ (See Figure~\ref{fig:allproblems}~(a)). 
Here, we show two possible scenarios when the temperature sensor sends $\mathtt{25^{\circ}C}$ and $\mathtt{35^{\circ}C}$ consecutively in a short time window.
In the first scenario, two measurements arrive at the cloud and are logged in the wrong order. The cloud then sends actuation commands based on the time the events are received. Thus, the commands arrive at the actuators in the wrong order.
In the second scenario, though the sensor events arrive at the cloud correctly, the commands arrive at actuators in the wrong order.
In both cases, the fan operates in a ``fan-on'' state rather than the desired ``fan-off'' state.

\vspace{1pt}\noindent\textbf{Single Actuator Invoked by Multiple Event Sources.} 
We consider different actuation commands given to a single actuator by multiple different event sources.
If different sensor events invoke the same actuation commands, the cloud may log the events in the incorrect order, yet the actuator's end state is consistent.
However, if the event sequence invokes different actuation commands, the actuator might receive it in the incorrect order. 
For instance, a user turns on and off a smart oven through two event sources, from a voice assistant and a mobile app (See Figure~\ref{fig:allproblems}~(b)).
The user first sends an oven-on command through voice and uses the mobile app to turn it off after a short time.
Thus, the user's desired state is ``oven-off''.
However, the events may arrive in the cloud in the wrong order causing the actuation commands received by actuators in the wrong order.
Although the events arrive at the cloud in the correct order, the delay between the cloud and oven may still change the order of the actuation commands. 
Both scenarios leave the oven in an undesired ``oven-on'' state.

\vspace{1pt}\noindent\textbf{Temporal/Logical Relations among Different Actuators.}
We study the order of actuation commands issued to multiple actuators, in contrast to a single actuator, invoked by single or multiple events. 
Incorrect ordering of actuation commands to multiple actuators leads to undesired states if there is a temporal/logical relationship between commands.
For instance, an app's ``close the sliding garage door'' and ``lock the garage door'' commands must be transmitted in the correct order (See Figure~\ref{fig:allproblems}~(c)). 
If the order changes, the garage door will attempt to lock before it is closed. This may leave the door opened or unlocked, creating unsafe conditions. 
Here, two separate actuators, a smart lock, and a smart garage door opener, are responsible for performing these actions.

\subsection{Formalizing the Misordering Problem} 
\label{sec:formalize}
We formalize each of the three issues introduced in Section~\ref{sec:misorder}.
Formal representation enables us to easily identify misordered events and actuation commands at each cloud and device (See Section~\ref{sec:Evaluation}).
%
We define the messages starting from an event and ending as an actuation command with the tuple $\mathtt{m = \langle s, e, a, c, ts, ta \rangle}$. An event is represented with a source $\mathtt{s}$ and an event $\mathtt{e}$, an actuation command is represented with an actuator $\mathtt{a}$ and a command $\mathtt{c}$, $\mathtt{ts}$ is the time the event physically occurs, and $\mathtt{ta}$ is the time actuation command is received at the actuator. Misordering problems occur when a sequence of messages are transmitted,  $\mathtt{\mathbf{M}= \{m_1, \ldots, m_n}\}$.

\vspace{1pt}\noindent\textbf{$\mathtt{P_1}$: Single Actuator Invoked by a Single Event Source.} 
$\mathtt{P_1}$ occurs when there exist at least two tuples $\mathtt{m_i}$ and $\mathtt{m_j}$ in $\mathtt{\mathbf{M}}$, such that they have the same event source and actuator, different actuation commands, and there is a discrepancy in the time sensor events are generated, and actuation commands are received. $\mathtt{P_1}$ is expressed as follows:
\vspace{-2.5pt}
\begin{equation*}
    \resizebox{1\hsize}{!}{$\mathtt{\exists m_i, m_j \in \mathbf{M}:~s_i = s_j \land a_i = a_j \land c_i \neq c_j \land ts_i > ts_j \land ta_i < ta_j}$}
\end{equation*}
%
\noindent\textbf{$\mathtt{P_2}$: Single Actuator Invoked by Multiple Event Sources.} $\mathtt{P_2}$ occurs when two tuples differ in event sources, have the same actuators with different actuation commands, and the order sensor events occur is different than the order actuation commands are received. $\mathtt{P_2}$ is expressed as follows:
\vspace{-2.5pt}
\begin{equation*}
    \resizebox{1\hsize}{!}{$\mathtt{\exists m_i, m_j \in \mathbf{M}:~s_i \neq s_j \land a_i = a_j \land c_i \neq c_j \land ts_i > ts_j \land ta_i < ta_j}$}
\end{equation*}
%
\noindent\textbf{$\mathtt{P_3}$: Temporal Relations among Different Actuators.} If two actuators have temporal relations, misordering causes problems regardless of the events. 
$\mathtt{P_3}$ is expressed as follows:
\vspace{-2.5pt}
\begin{equation*}
    \resizebox{0.7\hsize}{!}{$\mathtt{\exists m_i, m_j \in \mathbf{M}:~ a_i \neq a_j \land ts_i > ts_j \land ta_i < ta_j}$}
\end{equation*}
\section{Smart Home Deployment}
\label{sec:deployment}
\begin{table*}[t!]
\centering
\setlength{\tabcolsep}{-0.01em}
\renewcommand{\arraystretch}{1}
\caption{Descriptions of IoT apps installed in the house (Figure~\ref{fig:simulatedHome}).}
\resizebox{\textwidth}{!}{%
\begin{threeparttable}
\begin{tabular}{|c|c|c|c|c|}
\hline
\textbf{ID} & \textbf{App Description} & \textbf{Event} & \textbf{Actuation} & \textbf{Message Path} \\ \hline\hline
$\mathtt{M_1^*}$ & Start or stop the smart oven through the mobile application. & Icon Click & Smart Oven & Mobile App $\rightarrow$User IoT Cloud $\rightarrow$ Edge $\rightarrow$ Actuator \\ \hline 

$\mathtt{M_2^\dagger}$& Turn the smart plug on or off through the mobile application. & Icon Click & Smart Plug & Mobile App $\rightarrow$ User IoT Cloud $\rightarrow$ Edge $\rightarrow$ Actuator \\ \hline

\multirow{2}{*}{$\mathtt{M_3^\star}$}&\multirow{1}{*}{When an icon is clicked on the mobile application, first unlock then open } & \multirow{2}{*}{Icon Click} & Garage Door & Mobile App $\rightarrow$ User IoT Cloud $\rightarrow$ Edge $\rightarrow$ Actuator \\ \cline{4-5} 
& the garage door, or first close and then lock the garage door. &  & Garage Lock & Mobile App $\rightarrow$User IoT Cloud $\rightarrow$ Edge $\rightarrow$ Actuator \\ \hline

$\mathtt{M_4^\star}$& Close or open the window through the mobile application. & Icon Click & Window & Mobile App $\rightarrow$ User IoT Cloud $\rightarrow$ Edge  $\rightarrow$ Actuator \\ \hline

$\mathtt{TA_1^*}$ & Save periodic temperature sensor measurements to Google Spreadsheet. & Temp. Sensor & Google Spreadsheet & Sensor $\rightarrow$ Edge $\rightarrow$ User IoT Cloud $\rightarrow$ IFTTT Cloud $\rightarrow$ Google Cloud  \\ \hline 

\multirow{2}{*}{$\mathtt{TA_2^*}$} & \multirow{1}{*}{Activate the camera when there is a motion-active event,}  & \multirow{2}{*}{Motion Sensor} & \multirow{2}{*}{Smart Camera} & Sensor $\rightarrow$ Edge $\rightarrow$ User IoT Cloud $\rightarrow$ IFTTT Cloud $\rightarrow$  \\ 
 &  otherwise deactivate the camera. &  &  & User IoT Cloud $\rightarrow$ Edge $\rightarrow$ Actuator \\ \hline

\multirow{2}{*}{$\mathtt{TA_3^*}$} & \multirow{2}{*}{Stop the smart fan when temperature is below a user-specified threshold.} &  \multirow{2}{*}{Temp. Sensor} & \multirow{2}{*}{Smart Fan}   & Sensor $\rightarrow$ Edge  $\rightarrow$ User IoT Cloud $\rightarrow$ IFTTT Cloud $\rightarrow$ \\ 
&  &   &   &  User IoT Cloud $\rightarrow$ Edge  $\rightarrow$ Actuator \\ \hline

$\mathtt{TA_4^\dagger}$& Turn on the Hue light when the doorbell rings. & Doorbell & \multirow{1}{*}{Philips Hue Light} & Sensor $\rightarrow$ Edge  $\rightarrow$ User IoT Cloud $\rightarrow$ IFTTT Cloud $\rightarrow$ Hue Cloud $\rightarrow$ Actuator \\ \hline

$\mathtt{TA_5^\dagger}$& Turn on the Hue light when there is a motion-active event. & Motion Sensor & Philips Hue Light & Sensor $\rightarrow$ Edge  $\rightarrow$ User IoT Cloud $\rightarrow$ IFTTT Cloud $\rightarrow$ Hue Cloud $\rightarrow$ Actuator \\ \hline

$\mathtt{TA_6^{*\dagger}}$& Set or clear the thermostat through the IFTTT mobile application.  & Icon Click & Smart Thermostat & Mobile App $\rightarrow$ IFTTT Cloud $\rightarrow$User IoT Cloud $\rightarrow$ Edge  $\rightarrow$ Actuator \\ \hline

$\mathtt{IoT_1^*}$& Open window when smoke is detected, otherwise close the window. & Smoke Sensor & Window & Sensor $\rightarrow$ Edge  $\rightarrow$ User IoT Cloud $\rightarrow$ Edge  $\rightarrow$ Actuator \\ \hline

$\mathtt{IoT_2^*}$& Start the smart fan when temperature is above a user-specified threshold. & \multirow{1}{*}{Temp. Sensor} & \multirow{1}{*}{Smart Fan} & Sensor $\rightarrow$ Edge  $\rightarrow$ User IoT Cloud $\rightarrow$ Edge  $\rightarrow$ Actuator \\ \hline

\multirow{2}{*}{$\mathtt{IoT_3^\dagger}$} & \multirow{2}{*}{Activate the security alarm when all users leave the house, otherwise deactivate it. } & Presence Sensor & \multirow{2}{*}{Smart Alarm} & Sensor $\rightarrow$ Edge  $\rightarrow$User IoT Cloud $\rightarrow$ Edge  $\rightarrow$ Actuator  \\ \cline{3-3} \cline{5-5} 
 &  & Presence Sensor &  &Sensor $\rightarrow$ Edge  $\rightarrow$User IoT Cloud $\rightarrow$ Edge  $\rightarrow$ Actuator \\ \hline
 
$\mathtt{IoT_4^\dagger}$& Unlock the door when there is a motion-active event, otherwise lock it. & Motion Sensor & \multirow{1}{*}{Smart Lock} & Sensor $\rightarrow$ Edge  $\rightarrow$User IoT Cloud $\rightarrow$ Edge  $\rightarrow$ Actuator \\ \hline

$\mathtt{IoT_5^{*\dagger}}$& Lock or unlock the door through the voice-assistant. & Google Assistant & Smart Lock & Voice Assistant $\rightarrow$ Google Cloud$\rightarrow$ User IoT  Cloud $\rightarrow$ Edge  $\rightarrow$ Actuator \\ \hline

$\mathtt{IoT_6^\dagger}$ & Lock or unlock the door with a button click. & Button & Smart Lock & Sensor $\rightarrow$ Edge  $\rightarrow$ User IoT  Cloud $\rightarrow$ Edge  $\rightarrow$ Actuator \\ \hline

$\mathtt{IoT_7^{*\dagger}}$& Turn the Hue light on and off with a button click. & Button & \multirow{1}{*}{Philips Hue Light} & Sensor $\rightarrow$ Edge  $\rightarrow$User IoT Cloud $\rightarrow$ Hue Cloud $\rightarrow$ Actuator \\ \hline

$\mathtt{IoT_8^\dagger}$& When the door is open turn on the Hue light, otherwise turn off the Hue light. & Contact Sensor & Philips Hue Light & Sensor $\rightarrow$ Edge  $\rightarrow$ User IoT Cloud $\rightarrow$ Hue Cloud $\rightarrow$ Actuator \\ \hline

$\mathtt{IoT_9^\dagger}$& When power consumption exceeds a user-specified threshold, turn off the plug & Power Meter & \multirow{1}{*}{Smart Plug} & Sensor $\rightarrow$ Edge  $\rightarrow$User IoT Cloud $\rightarrow$ Edge  $\rightarrow$ Actuator \\ \hline

$\mathtt{IoT_{10}^\dagger}$ & When the window is open stop the thermostat, otherwise start the thermostat. & Contact Sensor & \multirow{1}{*}{Smart Thermostat} & Sensor$ \rightarrow$ Edge  $\rightarrow$User IoT Cloud $\rightarrow$ Edge  $\rightarrow$ Actuator \\ \hline

$\mathtt{IoT_{11}^\dagger}$& When there is a motion-active event, turn on the thermostat, otherwise turn it off. & Motion Sensor & Smart Thermostat & Sensor $\rightarrow$ Edge  $\rightarrow$ User IoT  Cloud $\rightarrow$ Edge  $\rightarrow$ Actuator \\ \hline
%

$\mathtt{IoT_{12}^\star}$ & Close or open the window shade through the switch. & Switch & Window Shade & Sensor $\rightarrow$ Edge  $\rightarrow$User IoT Cloud $\rightarrow$ Edge  $\rightarrow$ Actuator \\  \hline

\multirow{2}{*}{$\mathtt{IoT_{13}^\star}$} & Open the valve (take the moving sprinkler up or down) when the button is pushed, & Button & Buried Sprinkler & Sensor $\rightarrow$ Edge $\rightarrow$ User IoT  Cloud $\rightarrow$ IFTTT Cloud $\rightarrow$ IoT Cloud $\rightarrow$ Edge $\rightarrow$ Actuator \\ \cline{3-3} \cline{5-5}
 & start or stop the irrigation when the button is held.  & Button  & System & Sensor $\rightarrow$ Edge $\rightarrow$ User IoT  Cloud $\rightarrow$ IFTTT Cloud $\rightarrow$ IoT Cloud$\rightarrow$ Edge  $\rightarrow$ Actuator  \\ \hline

\end{tabular}
{\Large{$*$ indicates the app is used in Experiment 1, $\dagger$ indicates the app is used in Experiment 2, and $\star$ indicates the app is used in Experiment 3 in Section~\ref{sec:Evaluation}.}}
\end{threeparttable}
}
\label{tab:Apps}
\end{table*}

\begin{figure}[t!]
    \centering
    \includegraphics[width=\columnwidth]{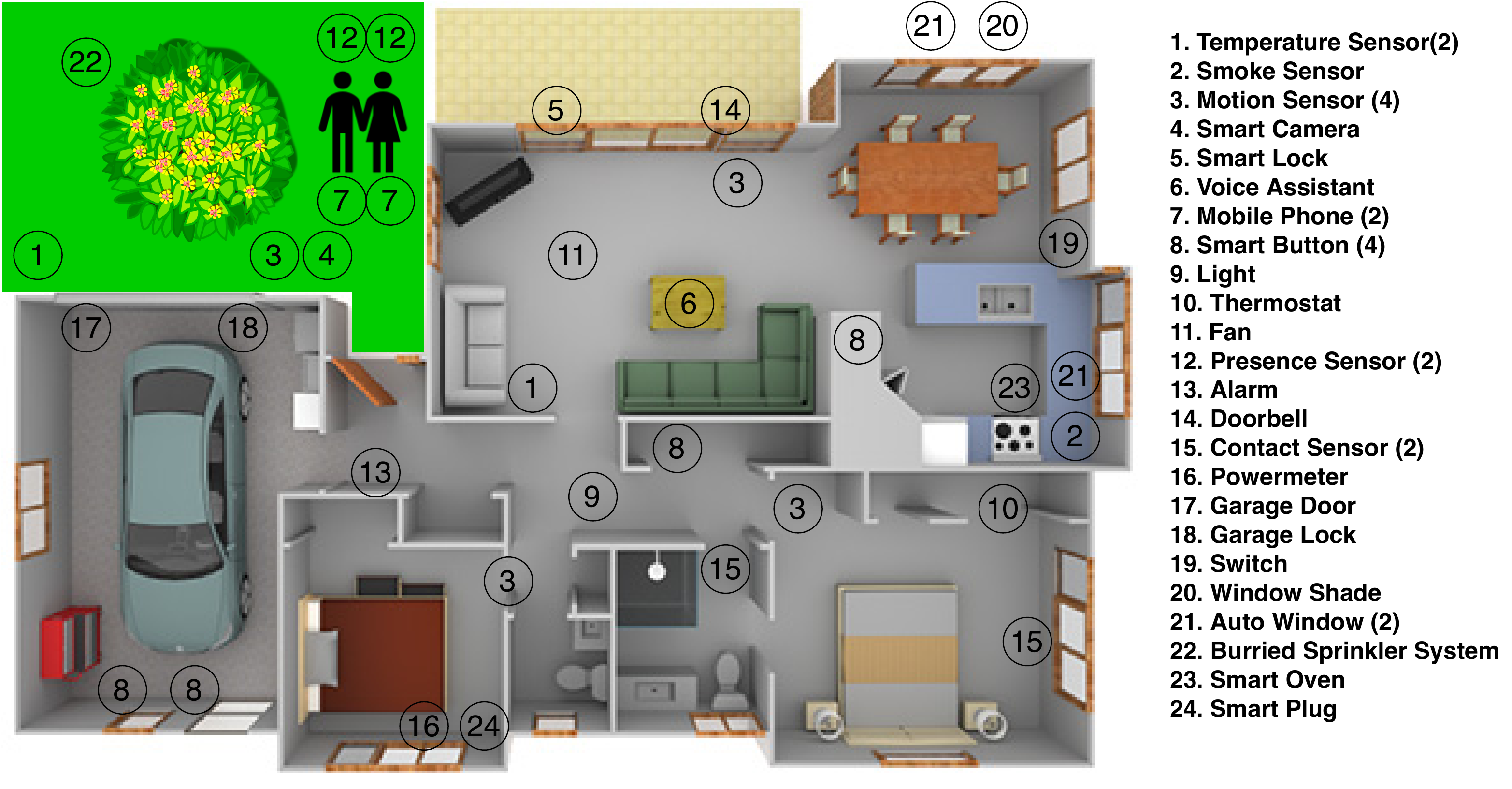}
    \caption{The simulated smart home used in our experiments.}
    \label{fig:simulatedHome}
\end{figure}

\vspace{1pt}\noindent\textbf{Testbed Environment.}
We evaluate the misordering problem in a simulated smart home containing $15$ different actuators and $7$ sensors, as a total of $35$ IoT devices (See Figure~\ref{fig:simulatedHome}).
We deploy $23$ apps to automate the devices. 
The apps include popular apps from IFTTT and SmartThings markets~\cite{iottestbench,celik2019iotguard}, and have diverse functionalities that encompass various real-life use cases.
Table~\ref{tab:Apps} presents the apps' IDs, descriptions, events, actuations, and communication paths from event creation to device actuation.
For instance, app $\mathtt{M_1}$ turns on and turns off the oven when the mobile app's icon is tapped. 
The path for $\mathtt{M_1}$ is mobile app $\rightarrow$ user IoT cloud $\rightarrow$ edge device $\rightarrow$ smart oven.
This enables us to capture different communication paths between different IoT components.

\vspace{1pt}\noindent\textbf{Implementation.}
We conduct the simulations in Ubuntu 20.04 OS running on 16 GB DDR3 RAM, 4 Processor cores. 
We convert the IFTTT rules and SmartThings apps to Python to work within the simulated home. 
Each edge device, cloud and IoT device is a separate process that communicates with each other over HTTP or the MQTT protocol.
The edge device communicates with the user IoT cloud and IoT devices through the Mosquitto MQTT broker V 1.6.8. IoT devices and clouds use the Python 3.0 Paho MQTT Client while the edge device is the MQTT Broker (server).
Vendor clouds and the actuators connected to these clouds (\eg Hue light) communicate over HTTP using the Python 3.0 HTTP servers module.
We set MQTT Quality of Service to $2$, the highest level, that ensures each message is received once. 
Additionally, we implement a logger to log sensor and actuator states in each component, used to evaluate the number of misordered events and commands.

\begin{table}[t!]
\addtolength{\tabcolsep}{4pt} 
\renewcommand{\arraystretch}{0.93}
\caption{Mean and standard deviations of processing time and network latency (secs) between IoT components.} 
\label{tab:delayTable}
{\footnotesize{
\centering
\resizebox{\columnwidth}{!}{
\begin{threeparttable}
\begin{tabular}{|c|c|}
\hline
 \textbf{Connection} & \textbf{Mean (sec)} $\pm$ \textbf{std (sec)} \\ \hline \hline
 IoT Devices $\leftrightarrow$ Edge Device &  0.056 $\pm$ 0.007 \\ \hline
 IoT Devices $\leftrightarrow$ Vendor Cloud & 1.4 $\pm$     0.3   \\ \hline
 Edge Device $\leftrightarrow$ User IoT Cloud & 1.5 $\pm$   0.4   \\ \hline
 Edge Device $\leftrightarrow$ Vendor Cloud & 1.5 $\pm$     0.4   \\ \hline
 Mobile App $\leftrightarrow$ User IoT Cloud & 1.5 $\pm$   0.4   \\ \hline
  Mobile App $\leftrightarrow$ Trigger-action Cloud & 2.5 $\pm$   0.5   \\ \hline
 User IoT Cloud $\leftrightarrow$ Trigger-action Cloud &2.5 $\pm$   0.5   \\ \hline
 User IoT Cloud $\leftrightarrow$ Vendor Cloud & 1.5 $\pm$  0.4   \\ \hline
\end{tabular}
\end{threeparttable}
}
}}
\end{table}

\vspace{1pt}\noindent\textbf{Delays among IoT Components.}
We use delays recorded among real IoT components from previous works~\cite{delayPaper, empiricalIFTTT}. These works study the response time of typical home IoT devices when multiple clouds interact with each other, such as user IoT and IFTTT clouds.
Table~\ref{tab:delayTable} shows the mean and std of delays in seconds.
The high and unreliable delays stem from ($1$) remote clouds that incur high round-trip time delays, ($2$) events traversing across multiple proprietary vendor clouds, ($3$) authentication overhead from TLS connections, and ($4$) the processing time overhead at the clouds due to the app and device identification~\cite{delayPaper}.
In our experiments, we randomly sample the delays from a Gaussian distribution using the mean and standard deviation of delays between each component.

\section{Evaluation}\label{sec:Evaluation}

We separately evaluate each type of misordering problem ($\mathtt{P_1, P_2, P_3}$) that we formally introduced in Section~\ref{sec:analysis} in 3 respective experiments.
In each experiment, we deploy a different set of apps in the house (Marked with $*, \dagger, \star$ in Table~\ref{tab:Apps}). 
We simulate the smart home when each app's event source generates $50$ consecutive events and over 8 different time intervals: $0.25, 0.5, \ldots, 2$ seconds.
These parameters are selected to conduct a stress test evaluating the misordering problem's extent under varying time intervals.
Finally, we analyze the misordered message (sensor events and actuation commands) rate at each component along the messages' path.

\begin{figure}[t!]
    \centering
    \includegraphics[width=\columnwidth]{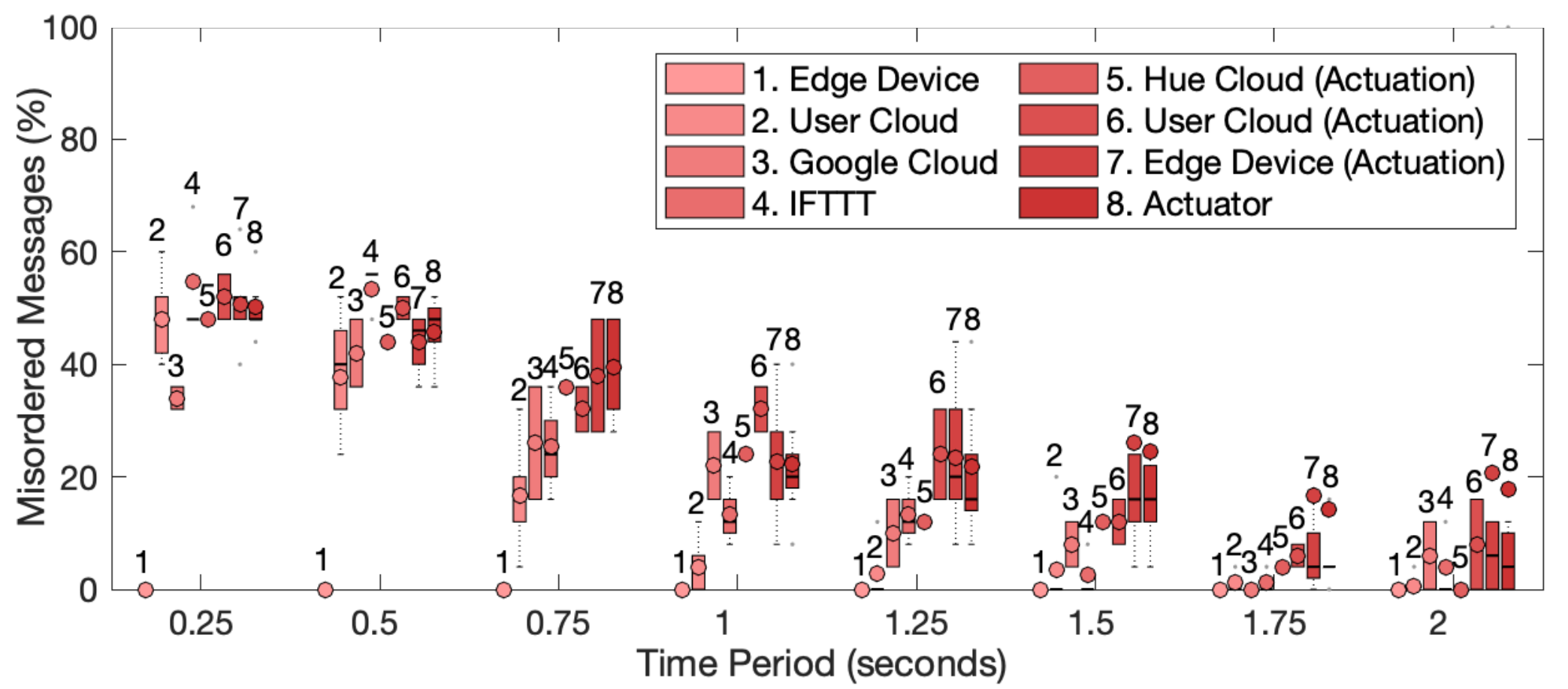}
    \caption{Exp. 1 - The misordered event and command percentage at each entity. Each bar represents min/max/mean/median of the percentages. Mean is marked ($\circ$) and median is marked (\textbf{--}).}
    \label{fig:exp1_NoSys}
\end{figure}

\vspace{1pt}\noindent\textbf{Experiment-1 ($\mathtt{P_1}$).}
We select $8$ apps from Table~\ref{tab:Apps} (Marked with $*$), each following different paths after an event occurs. 
We execute apps individually where consecutive events cause contradicting commands.
For instance, app $\mathtt{M_4}$ triggers $50$ consecutive {``open-window''} and {``close-window''} icon-click events (with different delays) from the mobile app.

Figure~\ref{fig:exp1_NoSys} presents the percentage of misordered messages that arrive at each entity in an app's communication path.
When the time between two consecutive messages (\ie time period) is $0.25$ seconds, on average, $48\%$ of the events arrive misordered to the user IoT cloud. 
This leads to an on average $50.3\%$ misordered actuation commands resulting in undesired system states. 
For instance, $48\%$ of the actuation commands from $\mathtt{IoT_5}$ (that controls the door's {lock} and {unlock} actuations) arrive to the actuator in an incorrect order. 
This causes discrepancies between the user's expected state (locked) and the true device state (unlocked), creating unsafe conditions with potentially disastrous consequences (\eg allowing a burglar to enter the home).
As expected, the rate of misordered messages decreases as the time between two consecutive messages increases.

\vspace{1pt}\noindent\textbf{Experiment-2 ($\mathtt{P_2}$).}
We identify the actuators that are controlled by multiple apps and conduct simulations with these apps (Marked with $\dagger$ in Table~\ref{tab:Apps}).
Particularly, we have identified five actuators, smart alarm, smart lock, Hue light, smart plug and smart thermostat, controlled by multiple apps.
In each simulation, event sources of these apps send out different consecutive events. 
For instance, $\mathtt{IoT_4}$, $\mathtt{IoT_5}$ and  $\mathtt{IoT_6}$ all control the door lock. Hence, we simulate these apps together, where an event of $\mathtt{IoT_4}$ is followed by an event of $\mathtt{IoT_5}$ and then $\mathtt{IoT_6}$, and each event is sent $50$ times.
The apps' events are selected such that they result in contradicting actuation commands, such as locking and unlocking the door.

\begin{figure}[t!]
    \centering
    \includegraphics[width=\columnwidth]{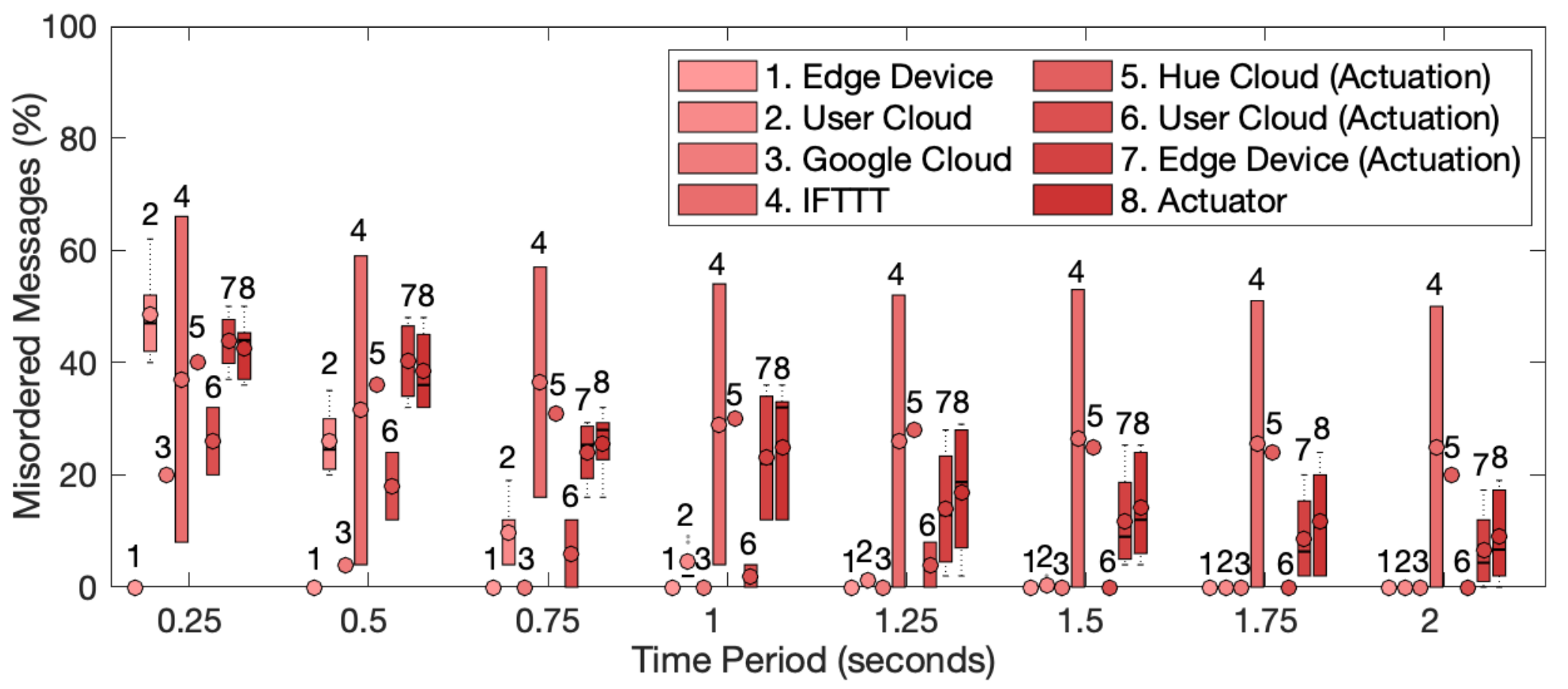}
    \caption{Exp. 2 - The misordered event and command percentage at each entity.}
    \label{fig:exp2_NoSys}
\end{figure}

Figure~\ref{fig:exp2_NoSys} shows the percentage of misordered messages with different time periods at each device and cloud.
On average, when the time period is $0.5$ seconds, $26.1\%$ of the events that arrive at the user cloud,  $31.5\%$ of the events that arrive at the IFTTT cloud, and $38.6\%$ of the commands that arrive at the actuators are misordered.
For instance, the door's smart lock may be left at an unexpected state due to the misorderings of the multiple events from $\mathtt{IoT_4}$, $\mathtt{IoT_5}$ and  $\mathtt{IoT_6}$.  
We observe that the misordered message percentage at the IFTTT cloud significantly differs. 
This is due to different paths an event may follow based on the apps. 
For instance, two of the apps that control the Hue Light do not include the IFTTT cloud on their paths, whereas the other two do. This causes additional delays for half of the events, resulting in a high number of misordered messages at the IFTTT cloud and the actuator.

\vspace{1pt}\noindent\textbf{Experiment-3 ($\mathtt{P_3}$).}
We have manually determined three groups of actuators with temporal relations:
($1$) The garage door should be closed first and then locked; and unlocked first and then opened (App $\mathtt{M_3}$). ($2$) The valve should be opened first, and then the irrigation should be started; and the irrigation should be stopped first, and then the valve should be closed (App $\mathtt{IoT_{13}}$). ($3$) The window shades should be opened first, and then windows should be opened; and the windows should be closed first, and then shades should be closed (Apps $\mathtt{M_4, IoT_{12}}$). 
We trigger events in the correct temporal order and check the percentage of misordered events and actuation commands.

\begin{figure}[t!]
    \centering
    \includegraphics[width=\columnwidth]{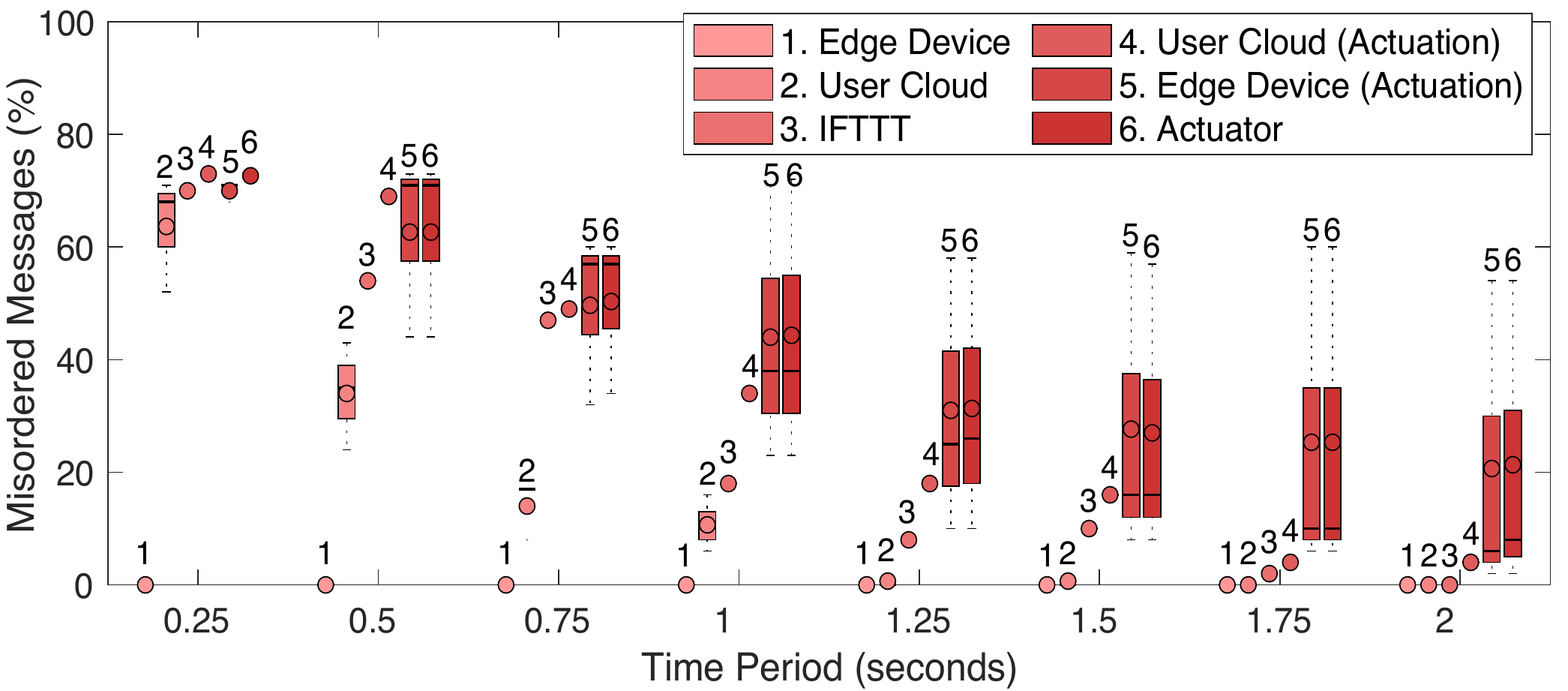}
    \caption{Exp. 3 - The misordered event and command percentage at each entity.}
    \label{fig:exp3_NoSys}
\end{figure}

Figure~\ref{fig:exp3_NoSys} presents the misordered event and command rates.
We observe the highest number of misordered messages at this experiment, compared to the initial two experiments.
This is because either the same event causes multiple actuations that must be received in a correct order or the paths that multiple actuators' actuation commands follow are different.
For instance, the {``icon-click''} event of $\mathtt{M_3}$ causes two actuations, opening/closing the garage door and locking/unlocking it. Since a single event causes two actuations simultaneously, $63\%$ of these commands are misordered on average over different time periods.  
These misordered messages cause safety issues when the garage door receives the lock command before close, since the lock may prevent the door from closing.

\section{Related Work}
Most attempts to achieve time-ordered events in distributed systems and wireless sensor networks (WSNs) capture event dependencies and ordering through Lamport timestamps~\cite{lamport1978ordering}, vector clocks~\cite{fidge1991logical}, and consensus-based approaches~\cite{schenato2011average}.
These systems mainly focus on a different problem that aims to manage dependencies by initiating schemes (\eg happens-before relation, vector timestamps) within each independent distributed system, and track them by monitoring the communication of the internal components. 
The TMOS system ensures sensor event ordering in WSNs by leveraging a group based protocol~\cite{temporalmessage}.
However, this cannot not be seamlessly applied to IoT since it relies on a globally synchronized clock among the devices, which is an open research problem in IoT~\cite{clockSync}. 
In distributed systems, more concern falls around total order involving multicast applications of publish/subscribe paradigms~\cite{distributed} and event dependency graphs~\cite{escriva2014kronos}. While their main concern is efficient and expected receipt of messages, these works do not account for maintaining the expected final system states. 
Additionally, there have been efforts to sort messages in event-time order to guarantee the correct event arrival for the edge and cloud~\cite{onishi2020recovery, correia2020omega, akidau2015dataflow, thein2014apache, wang2013query}. 
Yet, these approaches are not sufficient to guarantee the order of actuation commands' execution as they only ensure the event orders at the cloud side and not the resulting actuation of an event in a physical system.

\section{Conclusion}
As long as varied processing time and network latency continue to occur among different IoT components, misordered event sequences will occasionally occur.
Thus, with more devices and clouds working in conjunction, misordered event sequences may have compounding negative impacts resulting in  undesired system states.
In this paper, we thoroughly studied the misordering problem in centralized IoT systems. Our experiments showed that on average $29.8\%$ of the actuation commands are received in an incorrect order and they cause undesired system states.
Our findings encourage future work that respects the constraints of IoT devices to ensure correct event and actuation command ordering in IoT deployments.






\bibliographystyle{IEEEtran}
\bibliography{refs.bib}

%



\end{document}